\documentclass[acmsmall]{acmart}

\usepackage{caption}
\usepackage{subcaption}
\usepackage{multirow}
\usepackage{csquotes}
\usepackage{xcolor}
\usepackage{array}
\usepackage{longtable}
\usepackage{tabularx}

\AtBeginDocument{%
  \providecommand\BibTeX{{%
    \normalfont B\kern-0.5em{\scshape i\kern-0.25em b}\kern-0.8em\TeX}}}

\setcopyright{acmlicensed}
\acmJournal{PACMHCI}
\acmYear{2025} \acmVolume{9} \acmNumber{2} \acmArticle{CSCW176} \acmMonth{4}\acmDOI{10.1145/3711074}

\newcommand{\kp}[1]{``key-point#1''}
\newcommand{\ai}[1]{``action-item#1''}
\newcommand{\high}[1]{\textit{highlights}}
\newcommand{\hier}[1]{\textit{hierarchical}}
\newcommand{\chapter}[1]{\textit{chapter#1}}

\newcommand\leadin[1]{%
    \vskip 5pt \noindent\textbf{#1.} %
}
\renewcommand{\mkbegdispquote}[2]{\itshape}

\newcommand{\colortbl}{\color{black}}

\begin{document}
\title[Design, Implementation and Evaluation of an LLM-powered Meeting Recap System]{Summaries, Highlights, and Action Items: Design, Implementation and Evaluation of an LLM-powered Meeting Recap System}

\author{Sumit Asthana}
\email{asumit@umich.edu}
\orcid{0000-0002-8415-2508}
\affiliation{%
  \institution{University of Michigan, Ann Arbor}
  \city{Ann Arbor}
  \state{MI}
  \country{USA}
}

\author{Sagi Hilleli}
\email{sagih@microsoft.com}
\orcid{0009-0007-6359-0684}
\affiliation{%
  \institution{Microsoft}
  \city{Tel Aviv}
  \state{}
  \country{Israel}
}

\author{Pengcheng He}
\email{pengcheng.h@microsoft.com}
\orcid{0000-0002-1830-8893}
\affiliation{%
  \institution{Microsoft}
  \city{Redmond}
  \state{Washington}
  \country{USA}
}

\author{Aaron Halfaker}
\email{aaron.halfaker@microsoft.com}
\orcid{0000-0001-8907-6367}
\affiliation{%
  \institution{Microsoft}
  \city{Redmond}
  \state{Washington}
  \country{USA}
}

\renewcommand{\shortauthors}{Sumit Asthana, Sagi Hilleli, Pengcheng He, and Aaron Halfaker}



\begin{abstract}
Meetings play a critical infrastructural role in coordinating work. The recent surge of hybrid and remote meetings in computer-mediated spaces has led to new problems (e.g., more time spent in less engaging meetings) and new opportunities (e.g., automated transcription/captioning and recap support). Advances in dialogue summarization offer the potential for improving post-meeting experiences, but fixed-length summaries often fail to meet diverse needs, such as quick overviews or detailed insights. To address these gaps, we use cognitive science and discourse theories to conceptualize two recap designs: important highlights and a structured, hierarchical minutes view, targeting complementary recap needs. We operationalize these representations into high-fidelity prototypes using dialogue summarization. Finally, we evaluate the representations' effectiveness with seven users in the context of their work meetings at Microsoft. Our results show both recap types are valuable in different contexts, enabling collaboration through discussions and consensus-building. Exploring the meaning of users adding, editing, and deleting from recaps suggests varying alignment for using these actions to improve AI-recap. Our design implications, such as incorporating organizational artifacts (e.g., linking presentations) in recaps and personalizing context, advance the discourse of effective recap designs for organizational work and support past results from cognition studies.

\end{abstract}



\received{January 2024}
\received[revised]{July 2024}
\received[accepted]{October 2024}

\maketitle

\section{Introduction}
``Communication is the lifeblood of organizations''\cite{sethi2009interpersonal} and meetings are ``window into the soul of a business''\cite{karen2009effective} are truisms that describe the reality of modern work in organizational settings.  Meetings serve an important organizational purpose for people to discuss ideas, share information, build consensus, and make decisions. They help distribute key information, coordinate action, answer questions, and help people align their work efforts~\cite{whittaker2008design}. They also help reduce the organization's uncertainty by bringing the participants together to discuss and resolve issues~\cite{allen2015cambridge}. However, key details of the meetings could be missed or forgotten due to participants being oversubscribed to meetings or distractions by emails, IMs, and other communication~\cite{nathan2012}. Further, time conflicts and increasing cross-timezone collaborations cause participants to need to miss meetings altogether~\cite{nurmi2011coping}.

With the rise in geographically dispersed teams, the COVID-19 pandemic, and shifts towards remote work, organizations are increasingly adopting online web conferencing software to hold meetings\cite{NFW22, yankelovich2004}. The wide adoption of technically mediated meetings allows designers and researchers to apply technical approaches to support these new ways of working.

In this paper, we consider technological support for \emph{meeting recap} -- systems that aid in the capture of knowledge, decisions, and action items conveyed in a meeting for asynchronous review and engagement. A meeting recap is essential for preserving meeting content for attendees and non-attendees alike~\cite{whittaker2008design}. Traditional practices for capturing meeting content, such as paper and electronic note-taking, are often informal and burdensome on meeting participants~\cite{kalnikaitundefined2012}. 
Meeting content browsers provide a mix of audio, video, and ASR-generated transcripts with bookmark-style annotations~\cite{whittaker2008design,geyer2005towards} helping users in referencing information from recordings but do not eliminate the manual effort of searching and interpreting meeting content~\cite{arons1992techniques, moran1997ll}. This also limits asynchronous participation -- opportunities for non-attendees to engage with the meeting content and with attendees post-meeting -- due to the lack of context~\cite{kalnikaitundefined2012}. 

Recent advances in dialogue summarization in NLP~\cite{lewis2020} can generate contextual summaries from transcripts that can better support meeting recap goals. 
However, directly applying dialogue summarization on meeting transcripts to generate a fixed-length summary may not have the right structure and information to support diverse goals such as sharing knowledge, achieving consensus, negotiating deals, and deliberations~\cite{tuggener-etal-2021-summarizing,allen2015cambridge}.
For example, the same summary may be too verbose for participants needing quick highlights and too concise for those requiring in-depth context. Thus, an effective recap requires deciding what and how much should be summarized and what structure the recap should take to cater to these often diverging meeting goals. Unlike summaries of short dialogues, meetings have rich organizational context~\cite{whittaker2008design,kalnikaitundefined2008}. Hence, potential recap designs require expensive in-context evaluations of real meetings instead of inexpensive third-party crowdsourced evaluations~\cite{smith2020, star1994,schaekermann2018}. 

To address this gap in identifying effective recap designs and how they can support organizational work, we answer the following research questions.

\begin{itemize}
    \item RQ1: What \textit{key structures and information items} should be captured in a meeting recap to serve users' needs?
    \item RQ2: What \textit{benefits and challenges} do meeting attendees get for their \textit{personal and group work} from these key structures and information items in the meeting recap?
    \item RQ3: What do meeting attendees' post-meeting interactions with the recap mean for aligning AI-generated meeting recap with participants' needs?
\end{itemize}

\begin{figure}[h]
    \centering
    \includegraphics[trim=20 90 20 100,clip,width=0.9\linewidth]{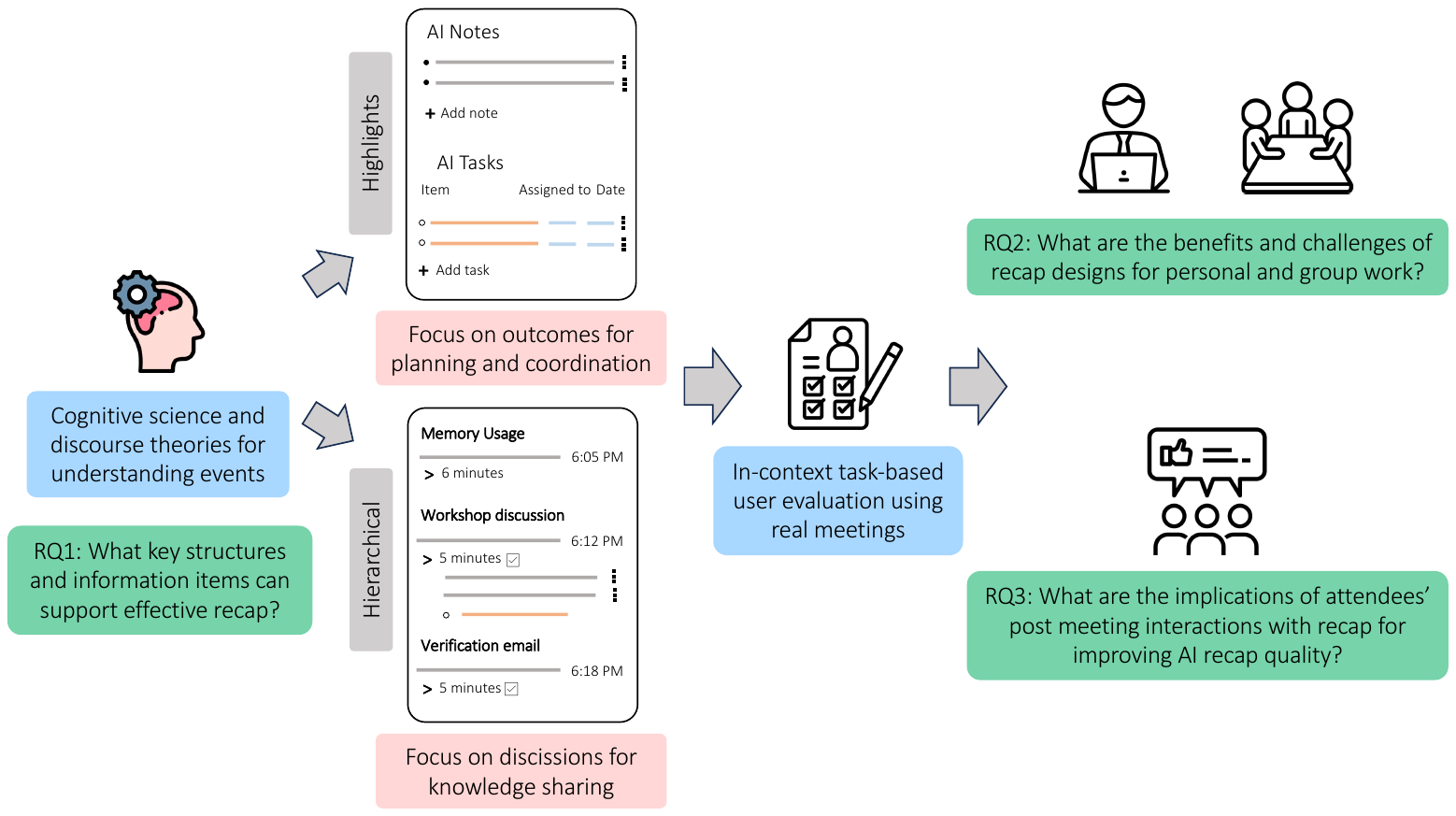}
    \caption{Overview of our work, highlighting the identification of salient recap designs to address information needs of meeting participants (RQ1), in-context task-based evaluation of the recap designs for personal and collaborative work using participants' real meetings (RQ2), and exploring implications of attendees' post meeting interactions with the recap for improving the AI recap (RQ3).}
    \label{fig:meeting_recap_overview}
\end{figure}

Building on CSCW scholarship of exploring technological support for complex work~\cite{knoll-etal-2022-user} through in-context evaluations of experimental high-fidelity designs, to answer our research questions, we 1) Use design rationales supported by theories from cognitive science and discourses~\cite{zacks2001perceiving} to conceptualize two \emph{recap designs}~\cite{yang2019} that serve diverse recap needs -- ``highlights'', that represent key moments in meetings for quick takeaways, and ``hierarchical'', that represents a meeting as chronological topic-focused discussions for sharing knowledge~\cite{kalnikaitundefined2008, nathan2012}, 2) Leverage the concept of \emph{experimental prototype experiences} from HCI~\cite{yang2019} to create high-fidelity operationalization our proposed recap designs using dialogue summarization, and 3) Perform preliminary task-based evaluations with seven information workers at Microsoft to explore the benefits and challenges when attendees use the recaps to support their work and collaborations. To address the scarcity of available meeting datasets for training~\cite{rennard-etal-2023-abstractive} and the need for personalization, we also seek to understand how user interactions could be used to improve the \emph{alignment}~\cite{yang2017role} of AI-generated recap models (see Figure~\ref{fig:meeting_recap_overview}).

Our in-context evaluations indicate that ``highlights'' and ``hierarchical'' representations serve complimentary recap needs. While both recaps provided comprehensible summaries, participants preferred highlights for quick overviews and sharing key decisions, and hierarchical for understanding discussions and sharing knowledge~\cite{geyer2005towards}. Participants preferred personalization in selecting relevant important notes for the highlights recap or personalized markers to support navigation in the hierarchical recap. 
Beyond personal work, participants' desire to use the recap as a collaborative document for consensus and knowledge work indicates the potential for meeting recap as an organizational artifact supporting articulation work~\cite{schmidt1992taking}. Participants indicated consistent reasons to add or edit summaries but many diverging reasons to delete summaries, indicating the nuances in user interactions for improving AI-generated summaries. These implications are grounded in the existing literature of increasing alignment of Human-AI models~\cite{yang2019clinic}. 

We contribute a (theoretically backed\cite{zacks2001perceiving, mann1987rhetorical,grosz-sidner-1986-attention}) foundation for \emph{designing meeting recap experiences} using automated summarization technologies that serve the practical needs of meeting attendees and non-attendees (examined in the context of work) and its role in supporting organizational work. Our work also discusses specific implications of using UX interactions with recap designs for aligning AI-recap with user's \emph{personal} recap needs.

\section{Related Work}
\subsection{Meeting Recap and Sensemaking}
Meetings reduce uncertainty, build mutual understanding, and help participants brainstorm new ideas~\cite{allen2015cambridge, acai2018getting}. A meeting recap sustains these benefits by documenting discussions and outputs, fostering collaboration and idea dissemination~\cite{bryant2005, kittur2007, andr2014, siangliulue2015}. Documented discussions serve as valuable resources for newcomers and experts, aiding memory and facilitating retrospective sensemaking that can lead to further insights or clarifications~\cite{zhang2018}.

Meetings both shape and are shaped by organizational culture~\cite{scott2023toward}. The value participants derive from meeting recaps depends on organizational expectations. Decision-driven meetings prioritize action items and key discussions, while collaboration-driven meetings focus on process discussions~\cite{scott2024}. Additionally, meeting discussions may not be linear, with participants using non-meeting tools for back-channel communication to resolve issues, seek clarifications, or improve decision-making~\cite{stephens2012multiple, dennis2010invisible}. Less structured meetings could also be non-linear, where participants visit previously discussed topics to seek clarifications or discuss further~\cite{park2024}.

Meeting goals and needs are also more importantly shaped by pre-meeting intentionality of participants, in addition to being shaped by organizational expectations~\cite{scott2024}. Adaptive interfaces that support efficient agenda planning during meetings~\cite{park2024} can lead to better expectation management for meetings. However, due to diverse roles and expectations from meetings, there is debate on what constitutes a good recap~\cite{whittaker2008design}. Including too much can overwhelm participants, leading to cognitive burden and fixation on ideas~\cite{moran1997ll, jansson1991design}. We propose two complementary meeting recaps: 1) focusing on key decisions and discussions to serve as ``cues'' for driving meeting agenda~\cite{kalnikaitundefined2008}, and 2) a ``minutes'' view providing a chronological record for knowledge work~\cite{whittaker2008design}. These approaches leverage cognitive fit theory~\cite{speier2006influence} and discourse interpretation theory~\cite{zacks2001perceiving,grosz-sidner-1986-attention,mann1987rhetorical} to align recap structure with these meeting goals. Our work on recap support is complementary to support for planning meeting goals. Better articulation of meeting goals also helps develop better recap experiences, which can inform the goal of future meetings. 

\subsection{Meeting Recap Tools}
Mediating work artifacts (e.g., design documents, action items, decisions) are critical in coordinating work activities~\cite{schmidt1992taking}.
People participate in meetings to create and modify these artifacts. Meeting recap tools can help record this activity for later reference through automatic transcription. However, due to high redundancy, raw audio and video recordings of meetings do not support sensemaking~\cite{moran1997ll, arons1992techniques}. Effective recap centers around making it easy to recall and modify these artifacts by marking timestamps in meetings, automatically extracting keyphrases for later edits, and combining meeting artifacts (e.g., slides, documents, whiteboard annotations, transcripts marked with discourse labels)~\cite{whittaker2008design}. Indexes into meetings or supporting them with short annotations are another popular way to help participants recall important moments~\cite{geyer2005towards, richter2003, whittaker1994, nathan2012,banerjee2005necessity}. For example, the Lite Minutes system~\cite{chiu2001liteminutes} allowed participants to create and share text, audio, and slide-based notes with automatically generated smart links and post-meeting note correction over email. When linked with video, such annotations provide recap value in learning environments as well~\cite{siangliulue2015, nathan2012} acting as digital bookmarks. Systems like Tivoli and Teamspace~\cite{geyer2005towards} directly support the organization and modification of artifacts (e.g., whiteboard diagrams, pen-written notes) in the meeting browser. While these tools provided affordances to manage meeting artifacts, participants could still face a high cognitive load due to multiple artifacts, and lack of interpretation of meetings.

To ease the cognitive load, meeting browsers also experimented with applying machine intelligence on transcripts, such as grouping semantically similar concepts to reduce redundancy~\cite{topkara2010tag, geyer2005towards, kazman1996, ehlen2008, manteidynamic, purver-etal-2006-unsupervised}, or automatically detecting action items~\cite{ehlen2007meeting}. Despite the rich modalities of recap browsers, Whittaker et al. ~\cite{whittaker2008design} found that participants found audio/video distracting and preferred scanning transcripts for their ease of skimming. However, this led to missing the high-level picture of meetings while being overconfident in their recall. Whittaker et al. and Banerjee et al.'s studies \cite{whittaker2008design, banerjee2005necessity} highlight the need for different recap forms: quick catch-ups and detailed, structured discussions. Previous recap systems\cite{whittaker1994, banerjee2005necessity, chiu2001liteminutes,ehlen2007meeting} allowed identification of low-level information but failed to provide a high-level narrative due to the limitations of text processing at the time. In the next section, we discuss the advances in NLP that we leverage in our system to enable the identification of salient information in meetings.  


\subsection{Speech and Text Summarization}
Speech and text summarization is the backbone of generating intelligent meeting recap experiences. Early meeting browsers used automatic speech recognition (ASR) to detect important meeting utterances directly. However, to leverage advances in natural language summarization methods~\cite{zhang-etal-2021-exploratory-study}, which are designed for text, it is more effective to use the ASR transcript of the meeting for recap generation~\cite{allahyari2017text, moratanch2017}. Recent neural-based summarization approaches~\cite{dong2018survey} can generate an abstract summary of the corpus with context. While corpus-based abstractive summarization provides more context and is useful for summarizing news articles or short dialogues, it does not work well for meetings due to long exchanges between multiple participants and topic shifts~\cite{khalifa-etal-2021-bag, chen2017, chen-etal-2021-dialogsum}. Directly processing large dialogue chunks to produce summaries can also lead to hallucinations even for the most advanced LLMs~\cite{tang2024tofueval}. 

To overcome the limitations of processing large dialogue chunks in a single pass, recent approaches have explored reducing the dialogue complexity by segmenting transcripts into topically focused areas and then summarizing the smaller segments~\cite{cohen-etal-2021-automatic,qi-etal-2021-improving-abstractive}. Another recent approach by ~\citet{hua-etal-2023-improving} explored representing the transcript in its abstract form using semantic relationships between concepts. Reducing the complexity of dialogues by breaking them into chunks can reduce hallucinations and provide better topic coherence; ASR errors in transcripts may introduce limitations such as the inability to detect multiple speakers~\cite {kanda2021investigation}. While identifying roles in multiple-speaker settings may complicate aspects like action items, converting ASR to text using text summarization still provides the best option to generate recaps for most meetings.   

On the evaluation side, exploring ideal designs for meeting summarization is complicated by the limitation of automated dialogue summarization metrics like ROGUE~\cite{lin-2004-rouge} to evaluate meeting recap goals reliably. This is because factors like the user's background, organizational contexts like established recap practices, and cognitive load can decide the user's contextual recap needs~\cite{whittaker2008design}. ~\citet{rennard-etal-2023-abstractive}'s survey of abstraction summarization specifically calls out the need for new metrics and datasets for evaluating meeting recap.

Thus, we hypothesized designs that reduce processing complexity for individual models by having each model focus on a specific aspect of processing (identifying \kp{s}, segmentation, summarization of segments)~\cite{cohen-etal-2021-automatic}. Our in-context evaluations by meeting participants themselves inform the design of effective recap and how participants use them for their work.

\section{System Design}
We now describe the design of our meeting recap system that helps us answer our three research questions. By informing the designs of our meeting recap system from cognitive science principles of how people interpret structured information in events, we answer the first research question.

\textbf{RQ1}: What \textit{key structures and information items} should be captured in a meeting recap to serve users' needs? 

\subsection{Design Rationale (RQ1)}
\label{sec:design_rationales}
Meeting recaps can support group work by allowing participants to recall important decisions, discussions, and knowledge from meetings and build on them~\cite{allen2015cambridge}. Identifying key structures and information items that support sensemaking for relevant recap needs (e.g., quick takeaways, a summary overview, and topics discussed) can reduce the cognitive burden for meeting attendees. We use cognitive fit theory to instrument design rationales\footnote{``Design rationales'' are the documented explanations of why specific design decisions were made during a product, system, or process development and how they were evaluated~\cite{lee1991s}.} that capture these key structures and information items relevant to the user's contextual recap needs~\cite{speier2006influence, lee1991s}. 

To answer \textit{\textbf{RQ1}: What key structures and information items should be captured in a meeting recap to serve users' needs?} We conceptualize the following two design rationales (DR1 and DR2).

\begin{figure}[h]
    \centering
    \begin{subfigure}[b]{.54\textwidth}
        \includegraphics[trim=0 0 0 0,clip,width=\linewidth]{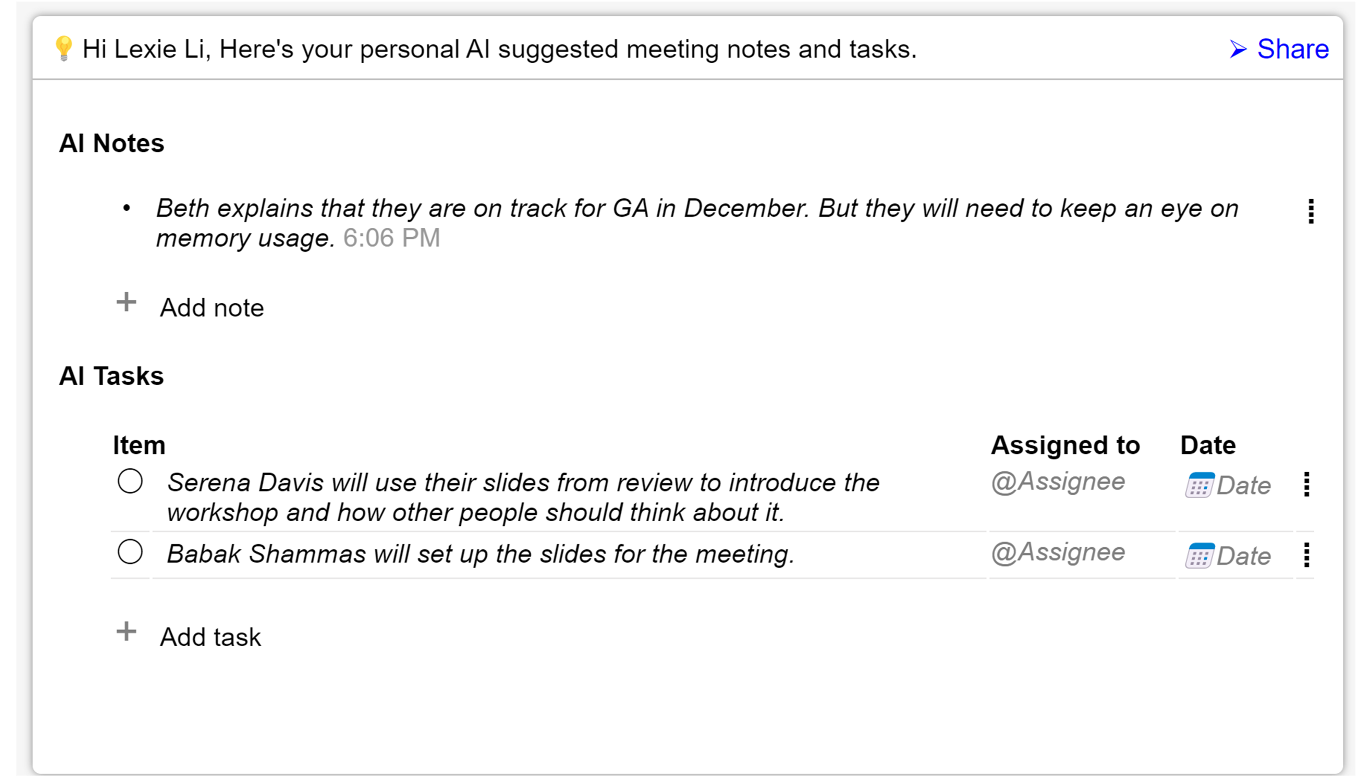}
    \caption{Highlights view}
    \label{fig:highlights}
    \end{subfigure}
    ~
    \begin{subfigure}[b]{.46\textwidth}
        \includegraphics[trim=0 0 0 0,clip,width=\linewidth]{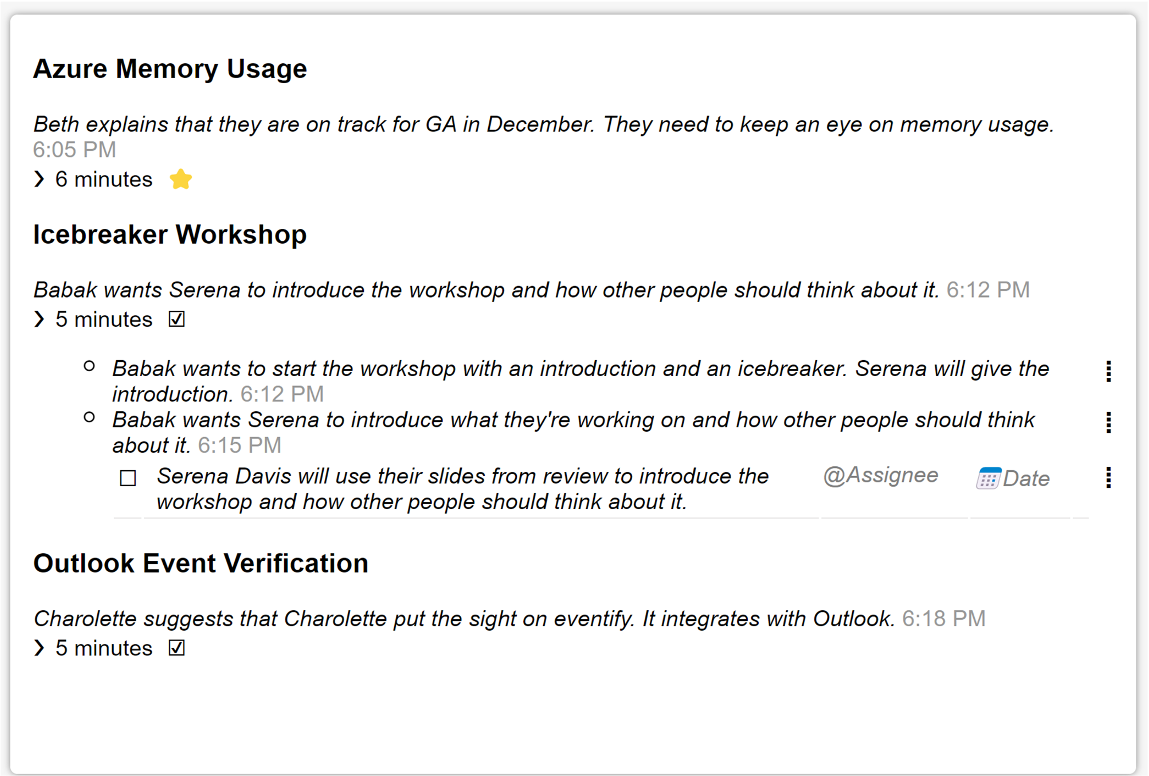}
    \caption{Hierarchical view}
    \label{fig:chapters}
    \end{subfigure}
    \caption{Meeting recap UX that exemplifies design rationales}
    \label{fig:experiences_ux}
\end{figure}

\leadin{\textbf{DR1 - Highlights: A meeting recap should be concise and focus on outcomes to support planning and coordination}} This rationale represents people's purpose for using meetings to coordinate their work, achieve consensus, and plan their upcoming tasks~\cite{allen2015cambridge}. The \emph{highlights recap} (DR1) focuses on pulling out key points and action items from meetings and representing them with one to two-sentence summaries. These are expected to help participants focus on the most important aspects of the meeting for decision-making without getting overwhelmed with meeting discussions (see Figure~\ref{fig:highlights}). 

    
\leadin{\textbf{DR2 - Hierarchical: A meeting recap should summarize the entire meeting, including discussions and outcomes within a hierarchical structure to support detailed context and knowledge sharing}} This rationale represents people's needs to understand discussions in meetings from a knowledge perspective and \emph{how} key decisions were taken. The \emph{hierarchical recap} (DR2) leverages a ``chapterization'' strategy to break the meeting into topically coherent ``chapters''.  Beneath each chapter, a user can explore lower-level summaries.  And under those summaries is the raw transcript.  Unlike the \emph{highlights recap}, the \emph{hierarchical recap} captures the entirety of the meeting in a recursive structure that resembles how people organize and interpret discourses~\cite{mann1987rhetorical} (see Figure~\ref{fig:chapters}).

\subsubsection{User interactions in each design for improving AI-generated recap}
In each recap, we also provided users affordances to add to, edit, and copy parts of the recap to share with others to understand how their feedback can help align the AI-recap with their expectations~\cite{christian2020alignment}. In the \high{} recap, we allowed users to \textit{add} notes and action items and \textit{edit and delete existing ones} to personalize their recap. To support planning and coordination, we also offered users the ability to \textit{assign tasks and due dates to users}. In the \hier{} recap, we allowed users to \textit{edit existing chapters} to personalize the chapters with their additional background knowledge. We also offered users the ability to \textit{mark notes within chapters} as \kp{s} or \ai{s} to increase the personal relevance of the recap and provide training data for models to learn about the aspects of the recap relevant to the user. These interactions were motivated by prior work that allowed participants to modify recap to improve their personal usefulness~\cite{geyer2005towards,gross2000towards} and team planning. We believe these interactions offer early indicators of the type and quality of training data we could gather at scale to enhance meeting recaps and AI through feedback. Detailed descriptions of the user experiences and interface components are in Section~\ref{sec:prototype}.

 

While prior recap studies~\cite{whittaker1994} have highlighted needs for both quick takeaways and detailed hierarchical discussions, we are the first to 1) Formalize the needs as design rationales to structure exploration, and provide supporting reasoning from cognitive science and discourse theories, 2) Design high-fidelity recap experiences based on the rationale to evaluate how well they meet recap needs of real meeting attendees in practice.

\subsection{Modeling}
We generate each recap using a set of transformer models. For each recap, we begin by processing the transcript utterances in the ASR text representation of the meeting discussion. We define an utterance as a sentence or a statement in the ASR representation of the transcript spoken as a continuous piece of speech without a pause~\footnote{https://en.wikipedia.org/wiki/Utterance}. 

\subsubsection{Highlights model}

We generate the highlights recap using four sequential transformer models~\cite{vaswani2017attention}: two for key points and two for action items. Each model pair consists of an extractive model (\emph{highlights\_extractive}) and an abstractive model (\emph{highlights\_abstractive})~\cite{cohen-etal-2021-automatic}. For each note or action item, the \emph{highlights\_extractive} model takes an utterance with its surrounding context as input and classifies it as a \kp{} or \ai{}. The \emph{highlights\_abstractive} model gets the utterances identified as a \kp{} or \ai{} from the \emph{highlights\_extractive} model and creates an abstractive summary of the utterance along with its surrounding context~\cite{cohen-etal-2021-automatic}.

The \emph{highlights\_extractive} model is a fine-tuned deBERTa with 12 transformer layers~\cite{he2020deberta}. It classifies utterances as key points or action items using a 106-token input size with context from surrounding utterances. This model was trained on ICSI and AMI labeled datasets~\cite{janin2003}. The \emph{highlights\_abstractive} model is a fine-tuned BART~\cite{lewis2020} model. It takes the output from the \emph{highlights\_extractive} model along with 512 context tokens and rephrases them in the third person. For example, if \emph{highlights\_extractive} identifies ``Serena: I will finish this by Friday'' as an action item, \emph{highlights\_abstractive} rewrites it as ``Serena will finish the slides by Friday.'' This approach makes the notes context-independent for independent review and sharing~\cite{whittaker2008design}.

Figure~\ref{fig:pipeline} (left) illustrates this pipeline. Please refer to~\citet{cohen-etal-2021-automatic}

\subsubsection{Hierarchical (``Chapters'') model}
\begin{figure}[h]
    \centering
    \includegraphics[trim={6cm 6cm 4cm 4cm,clip},width=0.75\linewidth]{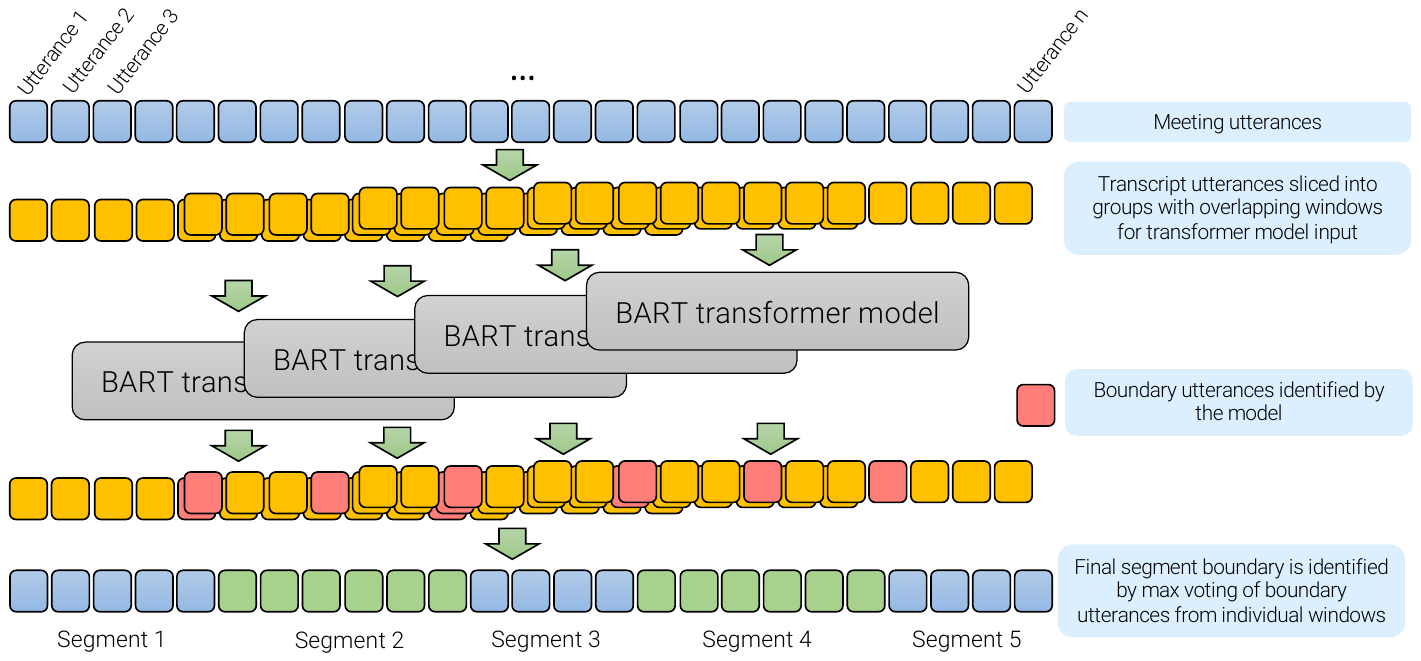}
    \caption{We apply the text-tiling segmentation process using BERT for segmenting long sequences of meeting dialogues into individual chapters (topics)~\cite{hearst-1997-text}. Utterance refers to a single sentence spoken by a meeting participant in the transcript.}
    \label{fig:segmentation-pipeline}
\end{figure}

We generate the hierarchical (``chapters'') recap in two steps -- 1) Segment the entire meeting transcript into segments or \chapter{s} where each \chapter{} corresponds to a sub-topic or set of topics (Figure~\ref{fig:segmentation-pipeline}), 2) Synthesize a title and a set of notes representing each \chapter{} (Figure~\ref{fig:pipeline} right half).

\textbf{Segmenting the transcript} We divide the meeting transcript into \chapter{s} using the \emph{hierarchical\_segment} model. \emph{Hierarchical\_segment} is a BART model that uses the text-tiling approach~\cite{hearst-1997-text, tur2011spoken}. The approach leverages lexical cohesion to determine segmentation boundaries where word distributions within segments are similar and across segments are dissimilar~\cite{galley-etal-2003-discourse, purver-etal-2006-unsupervised}. For prediction, we use a sliding window approach. In this approach, we break long input sequences into smaller windows of overlapping sequences. We then labeled each window with the classifier for topic boundaries, and the final boundary was identified with max-voting of the boundaries of the individual windows, resulting in segmented transcript blocks.

To train the \emph{hierarchical\_segment} model, we annotated a dataset of 12,600 meetings with chapter boundaries using crowd-sourced annotators from the UHRS platform\footnote{\url{https://prod.uhrs.playmsn.com/uhrs/}}, who marked transitions between topics. We split the dataset into 70\% training, 15\% validation, and 15\% testing. Annotators were compensated \$10 per transcript. We trained a BART~\cite{lewis2020} classification model on this annotated data to predict new segment starts. We split the transcripts into overlapping windows of 30 utterances with a stride of 10. We then applied the classifier to each utterance within these windows, and predictions were combined via maximum pooling to determine segment boundaries. This approach was necessary due to the token limit of 512 in transformer models~\cite{lewis2020}. The result was a transcript segmented into blocks, each representing a coherent topic.

\textbf{Generating Titles and Notes} For each of the blocks identified in the meeting transcript by the \emph{hierarchical\_segment} model, we generate its abstractive summary and title using the \emph{hierarchical\_abstractive} model. The \emph{hierarchical\_abstractive} model is a fine-tuned deBERTa model~\cite{he2020deberta}, trained on a dataset of 1M short (eight) dialogue utterances and their corresponding summary and rephrases them in the third person. The model generates notes, one for each sequential chunk of eight utterances in the meeting utterance blocks. We create the chapter headings using the \emph{hierarchical\_title} model. \emph{Hierarchical\_title} is a deBERTa model based on meeting utterances and topic assignments from another annotated dataset of 1M pairs.  This dataset has 1M meeting utterance topic assignment pairs and was also obtained through annotations on the UHRS crowd-work platform. 

\begin{figure}[h]
    \centering

    \begin{subfigure}[b]{\textwidth}
        \includegraphics[trim=350 300 130 450,clip,width=1.1\linewidth]{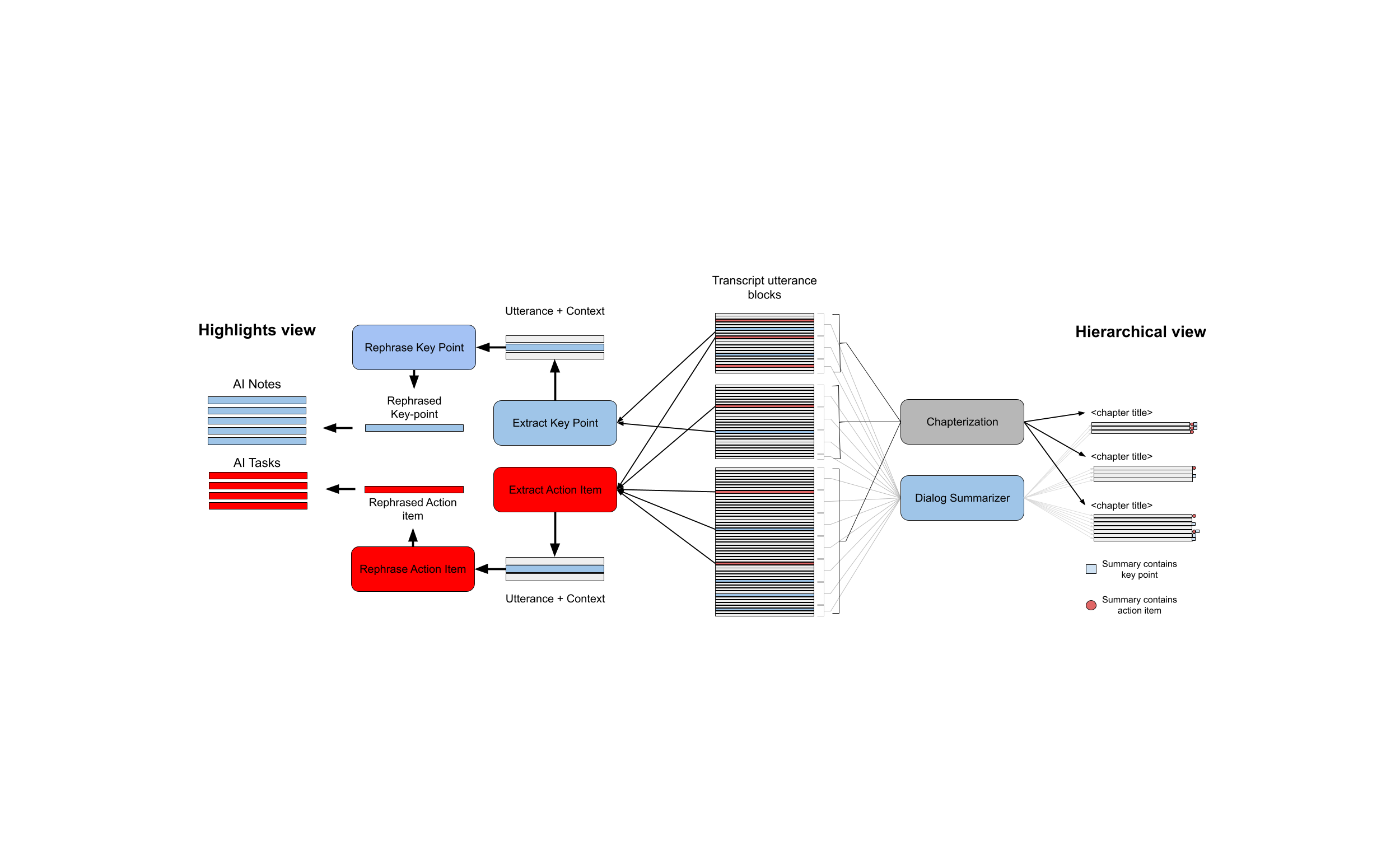}
    \caption{Summarization pipeline that generates the two recaps illustrated below. Highlights recap (left) extracts top-N utterances as key points or action items followed by rephrasing in 3rd person with context and displaying as a sequence (Figure~\ref{fig:highlights}). Hierarchical recap (right) segments the entire meeting transcript into sections, generates summaries for each section in 3rd person, and represents them sequentially in a minutes-like format (Figure~\ref{fig:chapters}). Summaries that contain key points or action items are marked with a corresponding number of stars and checkboxes, respectively. Each rectangular block is a transformer model. }
    \label{fig:pipeline}
    \end{subfigure}
    
    \caption{Illustration of the modeling pipeline that we use to generate the two recap experiences~(\ref{fig:pipeline}). 
    }
    \label{fig:experiences}
\end{figure}

The final output is a structured set of topics with headings and notes representing the meeting summary, like meeting minutes. Figure~\ref{fig:pipeline} (right) illustrates this pipeline. Table~\ref{tab:model_summary} describes all the models we used in our system for both recaps.

\begin{table}[h]
    \centering
    \begin{tabular}{|l|p{9.5cm}|}
        \toprule
        Model name & Description \\
        \midrule
        \textit{highlights\_extractive} & BART model that identifies the important utterances in the meeting around which the discourse needs to be captured in the highlights model (Figure~\ref{fig:pipeline} left).\\
        \textit{highlights\_abstractive} & deBERTa model that does an abstraction summarization of the context of the utterances identified by the highlights model (Figure~\ref{fig:pipeline} left).\\
        \textit{hierarchical\_segment} & BART model that identifies the segment boundary of transcripts that mark topic shifts (Figure~\ref{fig:pipeline} right)\\
        \textit{hierarchical\_abstractive} & deBERTa model that summarizes blocks of 8-utterances in each topic and represents them as a rolling summary of the meeting (Figure~\ref{fig:pipeline} right).\\
        \textit{hierarchical\_title} & deBERTa model that generates chapter titles (Figure~\ref{fig:pipeline} right).\\
        \bottomrule
    \end{tabular}
    \caption{Summary of models used in meeting recaps system design.}
    \label{tab:model_summary}
\end{table}



\subsection{Practical System Design Decisions}
Below, we summarize the rationales for the practical design decisions for our system implementation. We rationalize our choices using prior work and chose numbers that allowed for a reasonable system implementation. We use \citet{banerjee2005necessity}'s finding of 8 words per utterance and the widely accepted standard of 1 token = 0.75 words (or 1 word = 1.33 tokens) in NLP to explain the choices below.

\textbf{Input size of 106 tokens for the \textit{highlights\_extractive} model.}~\citet{cohen-etal-2021-automatic} identified that about ten utterance context is reasonable for detecting important highlights in the meeting. Thus, considering an average of 8 words per utterance~\cite{banerjee2005necessity}, our context window comprised about 106 tokens ($10*8*1.33$).

\textbf{Windows of 30 utterances and stride of 10 utterances for \textit{hierarchical\_segment} model} With 8 words per utterance~\cite{banerjee2005necessity}, a stride of 10 translated to a stride of about 133 tokens ($10*10*1.33$) and a window of 30 utterances translated to about 225 tokens ($10*30*1.33$). We leveraged ~\citet{koay-etal-2021-sliding}'s finding that a stride of 128 tokens and a window of 256 tokens yield the best performance.

\textbf{Surrounding context of 512 tokens for the \textit{highlights\_abstractive} and \textit{hierarchical-\_abstractive} models.}~\citet{he2022z} used a context window of 512 tokens to generate abstractive summaries for meeting related datasets in their experiments. They also identified that a similar context window is reasonable for identifying segment boundaries for topical segmentation of meetings. 

\textbf{Linearity of meetings} Our proposed design rationales work well for linear meetings but do not explicitly support non-linear meetings. Still, representing the meeting as chronological chapters in the \hier{} recap can support sensemaking for non-linear meetings as revisited topics will appear as chapters. Future work can augment the hierarchical approach by linking chapters across the meeting that pertain to the same topic, making it easier for participants to understand topic shifts to previous parts of the meeting.

\subsection{Prototype User Experience}
\label{sec:prototype}
We refer to \kp{s} as ``AI notes'' and \ai{s} as ``AI tasks'' in the UX for ease of understanding of participants. We use the shorthand terminology ``notes'' and ``tasks'' in the context of UX. To refer to both ``notes'' and ``tasks'' together, we use the term ``summary''.

We prototype the recap experiences on an HTML web app. The users start with a text area where they can copy-paste a transcript of a recorded meeting. Clicking ``Process'' sends the transcript to the backend to generate the recap insights for the two recaps. The backend uses the two summarization pipelines described above to create the recap data to send back to the web app front-end. Once the recap data is received from the two summarization pipelines, we populate the two tabs ``Highlights'' and ``Hierarchical'' with the data, formatting it as shown (Figure~\ref{fig:experiences}).

The highlights view displays a list of \kp{s} (``notes'') and \ai{s} (``tasks''). Key points are selected by the \emph{highlights extractive} model and rephrased by the \emph{highlights abstractive} model. Each action item includes editable ``assigned to'' and ``date'' fields for assignees and deadlines. Users can view the context for any item by clicking the three dots and selecting ``show context'', which reveals up to three transcript utterances before and after the detected note or task. Figure~\ref{fig:highlights} shows a screenshot of this UI.

The hierarchical view organizes the meeting by topics, each with a heading, a one-line summary, and a timespan. Clicking a timespan reveals a rolling summary with timestamps linked to transcript utterances. Key points and action items detected by the summarization algorithm are marked with stars and checkboxes, respectively, next to summaries and chapter titles. Clicking these markers expands the chapter, emphasizing the relevant summary.  Figure~\ref{fig:chapters} shows a screenshot of this UI.

To explore interactions for training data to align model behavior with user expectations, we implemented actions like add, edit, delete, and mark-important to refine summaries. In the highlights view, users could modify notes and action items, while in the hierarchical view, they could edit chapter summaries and titles and manage stars and checkboxes, though summaries couldn't be deleted. A ``share to chat'' feature enabled basic sharing. These affordances aimed to study user interactions to improve summary relevance, not to evaluate model feedback directly.

\section{User Study to Evaluate Meeting Recap Designs}
How attendees will use and benefit from the key structures and information items in the recap depends on their contextual needs, organizational culture, and their prior experience with technology~\cite{star1994, resnick1994, whittaker2008design}. Thus, we evaluate the effectiveness of our design rationales through task-based interviews about real meetings in the context of organizational work. We now describe our user study to evaluate the \high{} and \hier{} recaps within Microsoft workers' real meeting contexts. 




\subsection{Participants}
We evaluated the two recaps through semi-structured interviews with seven participants within Microsoft. We recruited the participants through internal lists and emails. When the participants agreed to the study, we set a 90-minute video conference interview time. Since we recruited employees within the organization as participants, we did not compensate them. 
\begin{table}[h]
    \centering
\begin{tabular}{ |l|l|l| }
\hline
\multicolumn{3}{|c|}{Demographic information} \\
\hline
\multirow{2}{*}{Gender} & Male & 50\% \\
 & Female & 50\% \\ \hline
 \multirow{5}{*}{Age} & 21-25 & 17\% \\
 & 26-30 & 50\% \\ 
 & 31-35 & -- \\ 
 & 36-40 & 17\% \\ 
 & 46-50 & 17\% \\ \hline
 \multirow{4}{*}{Region} & North America & 66\% \\
 & Greater China & 33\% \\ 
 & India & 0\% \\ 
 & Middle-east & 0\% \\  \hline
 \multirow{5}{*}{Profession} & Research & 58.3\% \\
 & Software engineer & 16.7\% \\ 
 & User engineer & 8.3\% \\ 
 & Design/creative & 8.3\% \\ 
 & Other & 8.3\% \\\hline
\end{tabular}
    \caption{Demographic details of interview participants}
    \label{tab:participant-demographics}
\end{table}

\subsection{Tasks and Procedures}
Two authors conducted each interview session: one leading and one taking notes. The lead interviewer explained the process, obtained consent to record, and used a recent meeting transcript to generate and discuss the meeting recap. We informed participants in advance to prepare a meeting recording, ensuring it was one to two weeks old. We began the sessions by asking participants to paste their meeting transcript into the prototype and start the recap generation. The system took some minutes to generate the entire recap. We started the interview by asking the participants preliminary questions about their meeting habits during this time. The prototype logged no user data, and meeting transcript contents remained private. During the interview, we asked the participants to share their screens to allow observation of their interactions with the recap.

We designed the interview questions to 1) Understand people's prior note-taking practices, 2) Understand the participants' \textit{actions and opinions} when attempting to use the recap for their personal and group work, and 3) Observe how participants interacted with the recap to understand how they would improve the recap and its implications for model development. Observing participants perform tasks instead of simply eliciting problems about the recap from them evokes higher-order cognitive capabilities~\cite{cotton2006reflecting} and is expected to provide data closer to their actual understanding of the recap. 

We divided the interview questions into three sections: 1) Participant background and meeting habits, 2) Tasks and questions for the highlights recap, and 3) Tasks and questions for the hierarchical recap. For both recaps, participants were asked to i) Identify important aspects for personal work, ii) Identify aspects for sharing, iii) Add missing summaries, iv) Fix inaccurate summaries, and v) Delete irrelevant summaries (if applicable).

We observed participants' actions, asked their reasons for choices, and encouraged them to think aloud while interacting with the UX to add, edit, or remove summaries. Participants explored the UX independently, with guidance only if they encountered difficulties. The think-aloud method provided insights into their thought processes and expectations~\cite{duncker1945problem}. The supplementary material provides the complete question set used for the interviews. 

\subsection{Analysis}
After each interview, one of the authors reviewed the recordings and transcripts and added any notes missed during the interviews. At the end of all the interviews, we had about 7 hours of recordings, associated transcripts, and notes taken during the interviews. We analyzed the interviews using thematic analysis~\cite{braun2006using}. 

\begin{table}[ht]
    \centering
    \colortbl \begin{tabular}{|p{0.8cm}|p{12.3cm}|}
        \toprule
        Parti-cipant & Meeting description\\
        \midrule
        1 & A 60-minute long research project meeting. The attendee was a research intern and the project lead. \\
        2 & A 45-minute long talk on a topic of Computer Architecture, with a 15 minute Q\&A. Attendees were one speaker from the domain of Computer Architecture and about 20 researchers in the same area.\\
        3 & A 60-minute long research update between a research mentor and a research intern.\\
        4 & A 60-minute research group meeting with the agenda of ``paper submission'' amongst two research mentors and a research intern.\\
        5 & A 60-minute kickoff meeting for a project between two teams, one research team and one product team. The agenda was to introduce collaboration between the research and the product team. The attendee was from the research team and attended the meeting while driving a car and could only partially pay attention to the meeting.\\
        6 & A 60-minute long weekly research update between a research mentor and a research intern. \\
        7 & A 60-minute long weekly research update. The discussion was on analyses and survey results. Two interns and their mentor participated in the meeting. \\
        \bottomrule
    \end{tabular}
    \caption{Overview of meetings}
    \label{tab:meeting_details}
\end{table}

\subsubsection{Coding}
We conducted a thematic analysis of the data. We used our design rationales (Section~\ref{sec:design_rationales}), research questions, and prior work~\cite{geyer2005towards} to define an initial set of codes focused on the strengths and weaknesses of the two recaps for supporting personal and group work. Thus, our initial codes captured the benefits and challenges of the recap designs, such as ``quick takeaways,'' ``detailed overview,'' ``supports decision-making,'' ``supports understanding discussions,'' ``helpful context,'' and ``lack of context.''

We started by applying these codes to the participants' responses about their recaps. Wherever applicable, we inductively refined our codebook with new open codes for flexible data exploration. For example, we observed participants indicate the need for ``personalization'' in both recaps and ``quick overview'' for \hier{} recap, which we had not identified before analysis. We also refined our initial codes, such as refining ``lack of context'' to ``lack of spoken context due to limited summaries'' or ``lack of external context'' such as videos or slides. We also coded participants' interactions with the UX to better understand how they adjusted the summaries to suit their preferences. Given the novelty of the LLM-powered meeting recap tool, we did not have initial codes for these interactions.

At the end of the initial open coding process, we had about 143 codes reflecting various recap dimensions that we wanted to study. We grouped the final codes into themes using affinity diagramming~\cite{holtzblatt2009contextual}. We categorized the codes based on their relevance to key aspects of meeting recaps, such as ``access to reminders,'' ``understanding discussions,'' ``access to context,'' ``personalization,'' ``achieving consensus,'' and ``sharing knowledge.'' Finally, we clustered related codes into high-level themes, such as ``benefits/challenges for personal work,'' ``benefits/challenges for group work,'' and ``user interaction with AI,'' representing our prominent findings in the paper.

Our team comprised of a research group and a product group where the research group preferred the \hier{} recap and the product group preferred the \high{} recap. The balanced preferences for both recap designs acted as a check on individual biases. Further, in weekly meetings, we challenged our assumptions about the designs to reduce bias in coding and aggregation. Our final results don't pick a favorite recap design but highlight the benefits and challenges of both recaps in different contexts, indicating our lack of bias towards any specific recap. 

\begin{table}[h]
    \centering
    \begin{tabular}{|p{3.7cm}|p{1.6cm}|p{4.3cm}|p{1.6cm}|}
        \toprule
        Construct & \#participants & Construct & \#participants\\
        \midrule
        Takes notes & 7 & Should take more notes& 2\\
        Keyword notes & 1 & Notes focus on ToDos& 4\\
        Digital notes & 7 & Recordings too long& 2\\
        Physical notes & 2 & Transcripts too long& 1\\
        Notes in chat & 2 & Notes help plan next meeting& 3\\
        Notes in Gdoc & 2 & Recaps with collaborator& 3\\
        Records meetings& 4 & Recaps with transcript& 3\\
        Shares notes& 6 & Recaps with recordings& 5\\
        Shares tasks& 2 & Recaps with slides& 1\\
        Shares slides& 0 & Recaps with chat& 2\\
        Cleanup before share& 1 & ASR issues& 1\\
        Collaborative notes& 4 & Task tracking& 4\\
        Agenda driven meeting& 1 & Recap at end of meeting& 1 \\
        Ask to record& 1 & & \\
        \bottomrule
    \end{tabular}
    \caption{Overview of meeting recap habits}
    \label{tab:general-meeting-habits}
\end{table}

\section{User Study Results}
We now describe the findings from our user study of the \high{} and \hier{} recaps. These findings answer our second and third research questions.

\textbf{RQ2}: What \textit{benefits and challenges} do meeting attendees get for their \textit{personal and group work} from these key structures and information items in recap?

\textbf{RQ3}: What do meeting attendees' post-meeting interactions with the recap mean for aligning AI-generated recap with participants' needs?


\renewcommand{\arraystretch}{1.3}
\begin{table}[ht]
\centering
\begin{tabular}{p{4cm}|p{4.7cm}|p{4.7cm}}
\toprule
 & \textbf{Highlights recap} & \textbf{Hierarchical recap}\\
\hline
\multicolumn{3}{l}{\textbf{RQ2}: Benefits and challenges for personal work} \\
\hline
\hline
\textbf{Access to quick reminders} & Provides quick reminders of key moments and action items for planning weekly tasks. & Easy to locate key moments and action items when catching up on the entire meeting but difficult when time is limited. \\
\textbf{Personalization of summaries} & Personalized selection of highlights could improve relevance. & Personalized relevance markers in chapters could support efficient navigation.\\
\textbf{Understanding discussions} & Summarizes key points efficiently but lacked depth in capturing different viewpoints. & Provides a comprehensive understanding of discussion nuances. \\
\textbf{Access to meeting context} & Provides key decisions but lacks context to understand their reasoning. & Requires effort to locate key decisions but helps understand the discussions' reasoning.\\
& & \\
\hline
\multicolumn{3}{l}{\textbf{RQ2}: Benefits and challenges for group work} \\
\hline
\hline
\textbf{Sharing decisions for consensus} & Quickly sending key decisions out in email/group chat helps ensure consensus and accountability. & Sharing chapters is less preferred for quick consensus and accountability due to verbosity.\\
\textbf{Sharing recap for brainstorming and knowledge sharing} & Sharing a list of key outcomes is not helpful for brainstorming. & Sharing chapters as collaborative notes could help brainstorm and provide clarifications.
\\
\hline
\multicolumn{3}{l}{\textbf{RQ3}: User interactions for model improvement} \\
\hline
\hline
\textbf{Consistent meaning for adding, editing summaries} & \multicolumn{2}{p{9.5cm}}{Additions or edits to summaries consistently meant that the resulting summary is more relevant to the participant.} \\
\textbf{Inconsistent meaning for deleting summaries} & \multicolumn{2}{p{9.5cm}}{Deleting summaries could mean less personal relevance, difficult to comprehend, or trivial item.}\\
\textbf{Ease of editing summaries} & Easy UX to edit summaries due to its clear and brief structure but limited context affected recall  & Full context around a decision or meeting point enabled easy edits.\\
\hline

\end{tabular}
\caption{Comparative summary of findings for highlights and hierarchical recaps for personal and group work and user-interaction implications for aligning AI recap with user's needs.}
\label{tab:summary_findings}
\end{table}

\subsection{Meeting Practices and Overview}

Table~\ref{tab:meeting_details} provides an overview of the meetings of our participants. Generally, all participants reported taking notes in one way or another during meetings.  All took some version of digital notes, while two participants also took physical notes with pen and paper. Four preferred putting notes into the chat or using a collaborative document format like Google Docs. Six participants also shared notes with others sometimes. Four participants suggested that their note-taking generally focused on To-Do's and task tracking. 

Participants reported several different strategies for recapping meetings: 1) Three asked someone who attended, 2) Three Used the transcript, 3) Five watched recording, 4) One indicated reviewing presented slides and 5) Two reviewed chat. These numbers are not mutually exclusive as each participant indicated multiple strategies for recap depending on their needs and time availability. Table~\ref{tab:general-meeting-habits} summarizes these general meeting habits across all participants.

Below, we describe the findings related to each of the recap designs. Table~\ref{tab:summary_findings} summarizes the findings described below across our two recap designs.


\subsection{Highlights Recap}
\subsubsection{Benefits and Challenges for Personal Use} \hfill

\leadin{Utilizing Highlights Recap as Quick Reminders} 
Four participants pointed out that suggested summaries served as reminders of what happened in the meeting.  Participants indicated using the summaries to decide their action items for the next week's meeting if it was a recurring one. They also said it helps them plan out their action items for the upcoming week if they have many action items and prioritize them. Some participants already understood what they needed to work on but still found the summaries helpful as \textit{reminders} (\texttt{P03}) to go back to when unsure. 

\begin{displayquote}
It is helpful to remember. They are accurate. Jogging my memory about the content of the meeting.  The task is helpful.  It reminds me that I can go look at the email about the meeting to see something I missed. -- \texttt{P03}
\end{displayquote}

\leadin{Inadequate Personalization in Note Prioritization} 
Six participants indicated that some of the generated highlights were not relevant to them. The model either did not capture notes from the relevant section of the meeting, or it captured action items that were low priority for the participant. For example, in a status update meeting, each participant discussed their own status sequentially, and the update given by others was not always relevant to the participant. To easily see discussions of relevant people, \texttt{P02} also suggested to \textit{group the summaries by people} so that they can pick the people relevant to them and see their discussions. Further, participants indicated that tasks assigned by their manager were high-priority action items that they needed to address urgently and termed it as ``main task'' other deliverables could be addressed later (\texttt{P03}). 

\begin{displayquote}
It's not that good because the AI tells you to do two things – one tiny task. A detail. Like, this detail is not correct, you have to fix after the meeting. It is not the main task. If these kinds of tasks are extracted, there will be plenty of to-do items, and I might miss the important ones. -- \texttt{P03}
\end{displayquote}

On finding less relevant summaries at the top, four participants also expressed the need to reorder them to have the most relevant items at the top for reviewing when in a hurry.


\leadin{Inadequacy in Capturing Comprehensive Discussions} 
Two participants indicated that discussions in the \emph{highlights} recap were not properly captured. We designed this recap to highlight key moments. 
For example, \texttt{P04} used ``show-context'' on an action-item to explore how the decision was made. The provided context of 3-utterances in this recap (Section~\ref{sec:prototype}) was not sufficient for them to understand the discussion around a paper submission.
\begin{displayquote}
We had a lot of discussion about what needed to be done. [...] If I were taking notes, I would have written down something to that effect. [...] Ultimately, we decided to work with the smaller dataset. No notes about any of that discussion. [...] It would have been nice to see notes around that. -- \texttt{P04}
\end{displayquote}

\leadin{Seeking Comprehensive Context in Summaries} 
Participants looked for more context to understand summaries better. In the specific instance of \texttt{P04}, part of their recap was a recorded presentation, and the context was a combination of utterances and visual information on the slides. After looking at the text-only summaries, \texttt{P04} remarked about access to the original video from the summaries.
\begin{displayquote}
If there was a way to link to the meeting for each of these notes, I could watch the relevant parts of the video based on the notes.  I can't scroll back in the context, so it would be great if I had the whole transcript in the context and a link to the video.  I may want some more detail.  If I can jump to the transcript or video, I can get that detail. -- \texttt{P04}
\end{displayquote} 

Participants also looked for context to understand who made the decision or why it was made. To get more context for ambiguous summaries, two participants explored the ``show context'' feature of the UX. However, three other participants still requested a link to the original video or the full transcript for more context. They mentioned that the limited context \emph{did not help them understand completely, and it would be helpful to look at the original discussion for clarity} (\texttt{P05}).  \texttt{P01} notes
\begin{displayquote}
I'm not sure what it is referring to and what it means in the context we were talking about. [...] (Participant expands context)  Oh! I remember what happened. A teams bug and the team couldn't see the slide deck. [...] So maybe 1st part got confused with 2nd part. Even for a human, it looks strange. -- \texttt{P01}
\end{displayquote}


\leadin{Addressing Pronoun Misassignments} 
Four participants indicated that the model generated summaries with their names associated with the wrong pronoun.  Participants had varying feelings towards such errors, ranging from attributing it as a \textit{minor point} (\texttt{P01}) to noticing and pointing them out or even feeling uncomfortable about it. This was more frequent with non-Western names, in which case technology's behavior induced feelings of non-inclusivity. 
\texttt{P06} reflected on the value of the summaries as well as the pronoun issues they saw.
\begin{displayquote}
Even just looking at the headers, I remember much more about the meeting at a glance.  The sentences have errors, and the pronouns are wrong. -- \texttt{P06} 
\end{displayquote}



\subsubsection{Benefits and Challenges for Group Work} \hfill

\leadin{Limited Utility for Collaborative Knowledge Sharing and Brainstorming} 
Due to the limited context in the \high{} recap, there was limited indication for use of the recap for brainstorming or sharing knowledge. Only one participant (\texttt{P07}) indicated using the \high{} recap to start a discussion around a \kp{} that the recap contained.

\leadin{Ensuring Consensus Through Collaborative Note Sharing}
Five participants wanted to edit the notes collaboratively for common visibility of changes. They viewed collaboration on the notes as a way to build consensus around high-value tasks and transparency and help identify dependencies between tasks. E.g., \texttt{P03} noted that collaboratively available notes are helpful to transparently identify if their task requires someone else to finish theirs first.

\begin{displayquote}
In meetings that involve more than two people, there are dependencies. e.g., someone else needs to address a task before I can do my task. It would be nice to have this dependency -- \texttt{P03}.
\end{displayquote}

\texttt{P05} imagined the \textit{shared summary document connected to people's organizational workflows} such as part of the meeting calendars so that it is easy for people to keep track of the summaries. Along similar lines, \texttt{P01} suggested that \textit{shared notes can also act as planning boards for teams, capturing team dependencies and next tasks for the team} (\texttt{P05}).




\begin{displayquote}
In meetings that involve more than two people, there are dependencies. e.g., someone else needs to address a task before I can do my task. It would be nice to have this dependency. -- \texttt{P05}
\end{displayquote}

For more accountability, (\texttt{P06}) further suggested \textit{to assign people in notes and notify via email to include people who might miss it otherwise}.

\subsubsection{Implications of User Interactions for Model Improvement} \hfill

\leadin{Consistency in Adding and Editing Highlights} 
All participants agreed that the intention behind adding and editing notes \& action-items was consistent (high alignment) in meaning. The notes that participants added were personally important to them~(4), a discussion to remember~(6), capture general topic/hierarchy~(4), or to add details to another note~(5). \texttt{P03} explained while adding a note from memory.
\begin{displayquote}
For the notes, I added them according to the timeline in my memory. In the beginning of the meeting, we discussed the result/trend. -- \texttt{P03}
\end{displayquote}

\texttt{P03} also reflected on adding notes that are important to them
\begin{displayquote}
I remember these because they are important parts I needed to do.  
[...]
I'm writing down the mistakes I made, the things I need to correct, and the main task. -- \texttt{P03}
\end{displayquote}


Regarding reasons for editing notes, two participants raised concerns about the AI's ability to learn from their edits because they used external knowledge to make the edits. Five participants edited the notes to add relevant context like ``\textit{<show on screen>}'' (\texttt{P01}) to refer to non-audio content. Three made major edits like rephrasing the note or task to make it \textit{more actionable} (\texttt{P03}). Four participants only changed the order of summaries like \textit{moving less relevant notes to the end} (\texttt{P01}) for easier review (\texttt{P01}). Participants indicated that the likelihood of editing notes depends on the quality. e.g., \texttt{P04} stated that they would edit the notes if the notes were of good quality.

\begin{displayquote}
It's easier to fix or clarify these mistakes than to start from scratch. If it takes less time for me to fix and correct stuff, then I'm going to do it. If its so bad that it takes more time, I'll probably stop using it. -- \texttt{P04}
\end{displayquote} 


\leadin{Inconsistencies in Deleting Highlights} 
Deleting notes and tasks was more complicated. Two participants reported that they might delete a task when it is done, three reported they might delete a task if it is redundant, and two others reported they might delete the task if it is inaccurate. 
One possible explanation for this is that it is not possible to recover deleted summaries, so participants were more wary of deleting than adding or editing, where the content can still be modified or reverted back. For example, \texttt{P03} reflected that \ai{s} which were already completed, wrong or redundant could be misleading.
\begin{displayquote}
If I see an inaccurate note or something I have already done, I will remove it. The wrong actionable could be misleading. I might delete redundant actions, insert a new task with the right meeting, and delete the task that was summarized wrongly. -- \texttt{P03}
\end{displayquote}

\texttt{P07}'s comment brings out the contrast between lack of a consistent meaning for deleting notes and a consistent meaning for adding or editing summaries 
\begin{displayquote}
For editing, I won't wait until I'm super confident.  

For delete, I'd be more careful.  I might delete something that is important to others.  If it learns from my deletions, I'm worried about that. -- \texttt{P07}
\end{displayquote}

\leadin{Difficulty in Editing Highlights Due to Incomplete Context} 
Four participants were able to add a note or a task from memory or using the context in UX, but three of these four needed to reference the raw transcript (that they had just copy-pasted into the prototype) in order to think of something they might add. Six participants were comfortable adding or editing summaries even if they were unsure or did not clearly remember the meeting details. 


\subsection{Hierarchical Recap}
\subsubsection{Benefits and Challenges for Personal Use}
\hfill
\leadin{Leveraging Hierarchical Recap for Comprehensive Overviews} 
One of the major reasons participants appreciated the hierarchical view was its ability to provide a quick overview of the meeting. While the \high{} recap offered fast access to reminders by only displaying the important meeting highlights, star icons in the \hier{} recap against chapters containing \kp{s} and \ai{s} also helped in easy access to reminders. Six participants indicated that a quick overview was useful to them. This overview helped a participant who \textit{missed the meeting and wants to understand in five minutes } (\texttt{P04}). Topic segmentation and progressive details helped participants easily navigate different parts of the meeting. It helped \texttt{P06} \textit{skip the poster session discussion and scan the analysis discussion} that they cared about. 
The hierarchical view was more intuitive, and \textit{looked more like a meeting to participants} (\texttt{P01}). \texttt{P05} was able to explore and understand the hierarchy on their own
\begin{displayquote}
This looks like a navigation tool which is more valuable.  [...]  This is two levels.  It seems right to me.  I see some duplication.  You have a main topic, a secondary topic, and then transcript. [...] This seems way more useful to me.  Even though there are errors, I can try to make sense of this. -- \texttt{P05}
\end{displayquote}

Three participants indicated errors in the chapters, but it did not affect their recall when they looked at the chapter headings and skimmed the chapter contents. This suggests that chapters helped participants recall meeting attendance beyond simply providing a record of what happened in the meeting.
\texttt{P07} said
\begin{displayquote}
I like the chapters or headers. Even just looking at the headers, I remember much more about the meeting at a glance. [...]
I like this more than [the highlights recap]. It gives more topics of what we discussed. [...]
[The highlights recap] is good for giving a rough understanding [...].   
I'd glance at [the highlights recap] and AI Tasks, then I'd go to the hierarchical and go into the sections we discussed.  It is easy for me to focus on the sections I care about. -- \texttt{P07}
\end{displayquote}

However, in alignment with the above quote, four participants indicated that glancing at \ai{s} in the \hier{} recap was still slower than the \high{} recap as the \hier{} recap required going over the entire meeting minutes.

\leadin{Moderate Personalization in Note Organization} 
Most participants did not have difficulty inferring checkboxes and stars as ``key points'' and ``action items'' respectively in the chapters. As participants navigated the \hier{} recap, it allowed them to favor navigation to chapters that contained more tasks quickly and assigned them to relevant people.
\begin{displayquote}
I'm choosing [this chapter] because it has a lot of action items.  [Picks an action item] This one I'd actually assign to [other person].  It's accurate.  He said it himself. -- \texttt{P03}
\end{displayquote}

However, similar to \high{}, \hier{} recap's \kp{s} and \ai{s} markers were not personalized. Participants indicated that personalized markers of relevant chapters could help them \textit{favor exploration of chapters} (\texttt{P05}) that were important for them, and skim other chapters. It is important to note that personalization in \high{s} recap related to selection of more relevant \high{} but personalization in \hier{} meant markers that favored personalization navigation.



\leadin{Comprehensive Discussion Understanding} 
All participants' manner of exploration of the hierarchical recap suggested a progressive, breadth-first exploration strategy. They would consider a relevant part of the meeting, locate its chapter heading (breadth scan), and explore the chapter by expanding it (depth scan), and looking at the notes that are part of the chapter.  If necessary, they would then expand a low-level summary to see the raw transcript that was being summarized. 

\texttt{P04}'s words exemplify this approach
\begin{displayquote}
I'm getting a summary flow of the meeting. [...] We had an attendee who missed a meeting recently.  I think they could understand this in 5 minutes.  I love the progressive detail. [...] I can understand this really quickly. -- \texttt{P04}
\end{displayquote}

Access to the full meeting context in a chapterized format allowed participants to explore the meeting in enough detail to support their recall. This contrasts with the \high{} recap, where participants found it difficult to understand discussions related to meeting highlights and decisions.

\leadin{Enhanced Comprehension through (Hierarchical) Meeting Context} 
The navigation provided by the \hier{} recap also allowed participants to get more context if they needed to delve deeper into the subject (see Section~\ref{sec:prototype}). Five participants used this context to expand discussions, understand who said what, and locate important notes and tasks. It helped \texttt{P02} understand the explanation in the Q\&A section of their meeting better.
\begin{displayquote}
\textit{(Picks a star) It explains the main difference between one project and another. Compared to other things, I'd say this is the main difference. Oh, this is important. (clicks the star bullet next to the other item) This is a very good question and answer, I don't know why this wasn't flagged at all.} -- \texttt{P02}
\end{displayquote}

However, similar to \high{}, absence of video access, or presentation materials limited sensemaking in scenarios when meeting discussions involved non-text artifacts (e.g., slides). \texttt{P02}'s attempt to understand a question related the Q\&A, where the speaker referred to a slide for part of an answer highlights the need for such context.




\subsubsection{Benefits and Challenges for Group Work} \hfill
\leadin{Effective Collaboration through Detailed Recaps} 
Five participants remarked that it was very easy to simply ``copy-paste'' a relevant meeting discussion into an email or on workplace chat to discuss with colleagues. This affordance was possible due to the context provided by chapters, where participants could open chapters to get more detailed summaries and choose the right level of details to share. \texttt{P06} remarked on how they could choose the right granularity to share the discussions.

\begin{displayquote}
For the first part of the meeting, I just wanted to confirm with my coworkers about the decision, but for the second part it was helpful to open up the details of the discussion and share it for further clarifications. -- \texttt{P06}
\end{displayquote}

Two participants also expressed the desire for follow-up discussions around the discussion minutes, similar to comments in a Google document, indicating the usefulness of recap for knowledge sharing and consensus building. Like the highlights recap, participants were more likely to contribute to collaboratively shared notes.

\leadin{Limited Utility for Consensus and Accountability} 
While the hierarchical recap helped participants get better meeting overviews, none indicated that they would use it to share their action items and decisions for consensus or accountability, as they did for the highlights experience. One possible reason is that the hierarchical recap contains meeting highlights as part of rolling summaries, and sharing discussions is too verbose for quick consensus and accountability. 

\subsubsection{Implications of User Interactions for Model Improvement} \hfill 
Participants indicated opinions similar to those in the highlights recap regarding adding and deleting notes in the hierarchical recap. However, they found editing notes relatively easy due to their ease of recall.

\leadin{Consistency in Adding and Editing Highlights} 
Five participants indicated that adding summaries to chapters or editing them held consistent meaning, suggesting they made changes to make the summaries more relevant. These participants were confident in their ability to edit the recap to fix any issues or add relevant context because of access to the full meeting context as chapters.

\texttt{P02} added more context from their background knowledge to a chapter note.

\begin{displayquote}
I am making this chapter summary more relevant for myself by taking note of when we started this discussion a month back. -- \texttt{P02}
\end{displayquote}

\leadin{Inconsistencies in Hiding Chapters} 
We did not allow participants to delete chapters as they represent the entire meeting. Instead, participants could adjust the depth of chapters they preferred in their recap by opening chapters they wanted more context on and hiding (collapsing) chapters they preferred less context on. We observed that participants reduced the depth of chapters for many reasons—errors in chapters and chapters that were less relevant to their work. This behavior is similar to participants' interactions with deleting highlights, which they did not prefer to be in their recap.


\begin{displayquote}
This second chapter has a lot of errors, and it is distracting to look at when trying to make sense of the meeting. -- \texttt{P06}
\end{displayquote}

\leadin{Ease of Editing Chapters Due to Summary Context} 
Four participants indicated that editing summaries in chapters was relatively easier than highlights because they had access to surrounding context in chronological order to aid their memory. Three even edited the summaries without difficulty or referring to the transcript, which participants had to resort to in the highlights recap because of insufficient context. 

\begin{displayquote}
I easily recalled what happened in this part of the meeting by looking at this chapter, so fixing this note was fairly easy! -- \texttt{P05}
\end{displayquote}

 


\section{Discussion}
In our design rationales, we hypothesized that recap designs that contain key structures and information items relevant to users' contextual recap needs would reduce the cognitive burden and facilitate recap. We identified two recap designs: one that focuses on key outcomes to support planning and coordination (\high{}) and another that creates an overview of the entire meeting within a hierarchical structure to enable knowledge sharing (\hier{}). Our user study indicates evidence that both recap designs reduce the cognitive burden for different recap needs. 

\subsection{DR1 - Highlights: Meeting Recap Should Focus on Key Outcomes to Support Planning and Coordination.} 
For the \emph{highlights recap}, we prioritized brevity and focus on meeting outcomes to support planning and coordination~\cite{moran1997}. Participants appreciated its value for reviewing attended meetings, particularly for quickly viewing tasks. However, they found it less helpful for understanding context, such as participants' opinions on topics or decision-making processes.

Determining the most relevant outcome to include in the highlights recap was a big issue for ML models. Some participants noted that the main points of discussion were not included in the suggested notes, while for some participants, the model generated no notes when it couldn't identify any utterance as important with high enough confidence. Some participants also suggested personalizing the order of the notes and tasks in their recap could help them process meeting takeaways faster. 

Our provided user interactions show potential for generating high-quality training data to better align~\cite{christian2020alignment} models with expectations. Participants found editing notes and tasks straightforward, aided by the ``show context'' option, and they reported high consistency in the meaning their edits conveyed. However, adding missing notes was challenging without revisiting the transcript, reducing the likelihood of addressing model omissions and limiting the informativeness for improving recall. Participants' concerns about deleting items highlight the need for more precise methods to indicate when an item should be excluded from highlights.

\subsection{DR2 - Hierarchical: Meeting Recap Should Summarize Discussions and Outcomes to Support Detailed Context and Knowledge Sharing.}
The design rationale for the hierarchical recap emphasized summarizing the entire meeting, including discussions and outcomes, in a structured hierarchical format inspired by prior work on how people understand discourses and temporal events~\cite{zacks2001perceiving,mann1987rhetorical}. Participants appreciated its alignment with their mental model of meetings, aiding quick overviews, knowledge sharing, and contextual understanding. The discourse-like structure made it easy for participants to search for relevant information despite occasional inaccuracies in chapter titles. A complete meeting summary helped contextualize key points and action items flagged by the system within the broader discussion.

Meeting recall supported by the \hier{} recap boosted participant confidence in editing summaries to better align with their needs due to the availability of relevant meeting context. While adding notes was less relevant for the hierarchical recap, selecting or deselecting highlights felt intuitive for participants. Participants were more inclined to correct highlights if the corrections benefited others or if they believed the AI would quickly learn from their corrections.

\subsection{Design Implications}
We now describe the design considerations for better meeting recap systems and designing scalable evaluations that applied to both our recap designs. Please refer to Table~\ref{tab:summary_findings} for a summary of the findings that inform these implications.

\leadin{Dialogue summarization supports recap sense-making} Contrary to Whittaker et al.~\cite{whittaker2008design} where participants found the summaries to be distracting and prevented high-level takeaways, all our participants found most summaries to be understandable and helpful when recalling parts of their meetings despite some imperfections. This allowed our participants to focus on higher-order needs with the recap, such as planning their upcoming tasks or the possible ways to share the summary to collaborate. Such observations were limited in prior studies where the cognitively demanding task of sense-making from transcripts prevented participants from thinking about how the recap could apply to their work~\cite{whittaker2008design, kalnikaitundefined2008}.

\leadin{Different meeting recaps for quick takeaways versus detailed discussions}
One of our key insights from comparing the \emph{highlights} and \emph{hierarchical} recaps was that they were both valuable in different contexts, complementing each other. 
While quick access to tasks was valuable to participants for planning after a meeting they attended, a rolling summary of the meeting helped get an overview if they missed the meeting or for a presentation. Nathan et al. ~\cite{nathan2012} and Whittaker et al. ~\cite{whittaker2008design} both hypothesized that a personal summary could act as a personal todo list, while a group summary could be helpful for collaborations and public contractual obligations. Recent work on meeting intentions by Scott et al.~\cite{scott2024} also indicates varying meeting needs that require different recap experiences. This exploratory finding also aligns with cognitive fit theory~\cite{speier2006influence} that the most useful representation of the summary will be the one that matches the structure of the task that the participants hope to achieve from the recap.

\leadin{Better context improves meeting recall}
All participants used the ability to expand the ``context'' of a summary into the original dialog to make sense of a suggested note, suggested task, or a low-level summary (see Figure~\ref{fig:experiences_ux} for context affordance).  Many participants also requested deep links to video recordings to review the discussion quickly. This aligns with prior findings~\cite{geyer2005towards, moran1998, whittaker1994} showing that timestamped audio or verbatim speech boosts recall and aids note correction. One participant referenced non-speech content, highlighting the value of integrating dialogue summaries with artifacts like slides and files for a holistic meeting view~\cite{topkara2010tag, cruz1994capturing, he1999}.
\leadin{Meeting recap as a natural artifact for collaboration}
Prior meeting recap systems~\cite{ehlen2007meeting, banerjee2005necessity} focused on the effective individual recap, but their collaboration aspect was under-explored due to a lack of mature collaborative editing technologies. As an exception, Geyer et al.~\cite{geyer2005towards} designed a customized system to associate domain-specific artifacts with indexes in meetings for later collaboration. However, their evaluations highlight the difficulty of using the recap system's artifacts outside the system because the recap is part of a highly customized interface.



Our system generates a recap resembling manual note-taking, similar to collaborative documents like Google Docs or Microsoft Word~\cite{samuli2003}. Participants expressed interest in using the recap for post-meeting work, such as clarifications, follow-up discussions, and sharing key decisions. This collaborative recap supports remote work and cross-timezone collaboration, enabling asynchronous participation by consuming the recap and continuing discussions~\cite{richter2001}.

\leadin{Permanence and transparency of recap can change work practices} 
Meeting recaps foster a collaborative workspace, enhancing transparency and potentially changing work practices~\cite{smith2020}. In the absence of recap, there is limited post-meeting agreement and risks of forgotten or fictitious ideas~\cite{mantei1988, olson1992small}. Participants found the \high{} recap valuable for consensus, transparency, and accountability, even for non-attendees. By making discussions explicit, permanent, and shareable, meeting recaps could shift work obligations, similar to how algorithmic support impacted Wikipedia practices~\cite{halfaker2016ores,kling1992cscwprivacy,scott2023toward}. Given the impact of meeting recap in shaping personal and collaborative work boundaries,  future work should explore how recap affordances can be consentfully integrated into work practices.


\begin{table}[h]
\centering
\colortbl \begin{tabular}{|p{6cm}|p{7cm}|}
\toprule
User behavior & Implication for system \\
\midrule
User edits summary & Summary item after the edit is better quality \\
User shares summary & Summary item is important to user \\
User opens a section in hierarchical view & Section is relevant to the user \\
User looks up source dialogues for summary & Summary item is relevant to the user, and possibly lacks full context\\
User deletes summary & Summary could be non-relevant, wrong or poorly written\\
\bottomrule
\end{tabular}
\caption{User interactions with the UX that can be used to provide feedback signals to the model on summary quality.}
\label{tab:d2l}
\end{table}

\leadin{Designing to learn}
The models for recap experiences were trained on crowd-sourced data, where workers summarized and labeled \kp{s} and \ai{s} from meetings they did not attend. This creates a gap between the understanding of attendees, who are part of the social context, and labelers, who only read transcripts~\cite{schober1989understanding}. 
Consequently, AI systems trained on crowd data often produce generalist models that work well in low-context scenarios but struggle in scenarios that involve understanding users' contextual needs~\cite{yang2017role}. Additionally, privacy concerns and lack of context limit available meeting datasets~\cite{rennard-etal-2023-abstractive}.

Thus, for such applications, it is important to identify methods to improve systems through in-context user interactions~\cite{ehlen2008}. However, these interactions may not always have consistent meaning (alignment). In our study, participants showed consistent alignment in adding and editing notes and tasks, indicating additions and edits could be high-quality feedback signals for AI to learn from. Most participants were more likely to edit if it benefited others or for group sharing. However, deleting notes was more nuanced, with varied reasons such as irrelevance to tasks, lack of clarity, or low relevance, indicating the need for careful consideration when using deletion of notes as feedback for AI.


Therefore, when designing AI to learn from user actions, it is important to understand the reasons behind those actions. Feedback from additions and edits can help retrain models to improve AI performance. However, feedback on incorrect or deleted notes or tasks should be collected in a way that clarifies user intentions. Table~\ref{tab:d2l} summarizes user behaviors and their implications for system improvement.

\section{Limitations} 
\textbf{Study limitations.}
Our limited and homogeneous participant pool places some limits on the generalization of findings. All our participants had a basic technical background, limiting the exploration of recap use for diverse user backgrounds. Further, participants' familiarity with technology might have influenced their acceptance of minor AI errors (e.g., minor ASR errors). Yet we observed that this familiarity did not prevent participants from identifying important issues such as pronoun mistakes. Regarding organization, other organization contexts (e.g., financial, legal) may find novel uses for meeting recaps that we did not cover and encourage exploration in future studies. 

An important requirement for this study was participants needed to pre-record meetings. Recording was not commonly accepted due to privacy concerns and norms around the persistence of discussions. Some participants expressed some discomfort when asked to record meetings but complied after clarifying the protection of their data under the study protocol. Future research should consider how to manage privacy aspects when studying meeting recaps.  

\textbf{System design limitations.} 
Our design focuses on identifying key meeting moments (\high{}) or providing a rolling summary (\hier{}) as meeting minutes. Some meetings involve revisiting earlier topics~\cite{park2024}. \textit{Hierarchical} recap does not break the pattern of non-linearity, as revisiting prior topics after new topics are introduced will still be captured as chapters. To enhance this, future work can explore additions of explicit text or visual references that link related discussions, helping participants understand the non-linear structure. Our system uses ASR transcript as input for summarization, which could attribute the same name to multiple speakers if they attend from the conference room. In such scenarios, additional intelligence is needed on the ASR side to distinguish speakers based on voice characteristics (e.g., tone).

We do not employ state-of-the-art LLMs for our work; instead, we use a sequence of low-resource transformer models that we trained on summarization datasets. The sensitivity of participants' real organizational meeting data motivated this choice. Because our system generates meeting recap in a two-step process, instead of a single model processing the entire transcript, our system divides the complexity of the task across several transformer models (e.g., identifying \kp{s} and \ai{s}, abstract summarization of small dialogue segments)~\cite{rennard-etal-2023-abstractive}. Larger models like GPT4 also introduce errors in abstraction summarization~\cite{ramprasad2024analyzing}. However, such models may provide additional benefits such as ``circumstantial inference'' where the summary is not directly stated in the meeting but inferred from the dialogues~\cite{ramprasad2024analyzing}. Future work could explore how meeting recaps can benefit from such inferred summaries. 



\section{Conclusion and Future work}
As meetings increasingly shift to hybrid or online formats, effective recaps can help participants manage their meeting workload. We propose a meeting recap system with two types: 1) a quick highlights recap and 2) full meeting minutes segmented into chapters. Our task-based in-context evaluation with participants' real meetings reveals that participants value both recaps. Organizational size, processes, and culture may influence recap benefits. Future work can expand on these findings to evaluate recaps across various organizational and cultural contexts.

Designing effective meeting recaps is challenging as ML models trained on crowdsourced data often do not align with target audiences' needs (e.g., organization workers)~\cite{yang2019, star1994}. This highlights the need for interfaces that generate useful training data through ``natural'' use~\cite{yang2017role}. Our findings suggest designs that let participants interact with notes to reflect their preferences, such as quick edit suggestions or memory aids. Future work can conduct quantitative experiments to validate the quality of training data gathered from user interactions in meeting recap applications.


Our findings suggest expanding meeting notes beyond recaps to facilitate discussions that influence ideas, consensus, and information exchange. Integrating highlights or hierarchical recaps in tools like Microsoft Teams or Slack could improve understanding of work processes. Designers must balance privacy, visibility, and team dynamics when incorporating meeting recaps into corporate tools. Addressing concerns about recording meetings and exploring privacy-preserving solutions, where participants control the recap content, is an important avenue for future work.

\begin{acks}
We thank the team at Office of Applied Research at Microsoft for their feedback in improving the work. We also thank the participants at Microsoft who provided valuable feedback about the usefulness of the recap designs in the context of their real work meetings. 
\end{acks}

\bibliographystyle{ACM-Reference-Format}
\bibliography{sample-base}

\appendix









\end{document}